\documentclass[a4paper, amsfonts, amssymb, amsmath, reprint, showkeys,  twoside,twocolumn,superscript address]{revtex4-2}

\usepackage{graphicx}
\usepackage{graphics}
\usepackage{amsmath,amssymb}
\usepackage[usenames]{color}
\usepackage{subfigure}
\usepackage{hyperref}
\usepackage{breakurl}
\usepackage{enumitem}
\usepackage{lgreek}
\usepackage{enumitem}

\usepackage{bm,amssymb}

\renewcommand\Re{\operatorname{Re}}
\renewcommand\Im{\operatorname{Im}}

\newcommand{\be}{\begin{equation}}
\newcommand{\ee}{\end{equation}}
\newcommand{\la}{\langle}
\newcommand{\ra}{\rangle}

\newcommand{\mez}{\hspace*{+0.50cm}}

\newcommand{\m}{\hspace*{-0.50mm}}
\newcommand{\n}{\hspace*{-0.25mm}}

\renewcommand{\Re}[1]{\hbox{Re} ~ #1}
\renewcommand{\Im}[1]{\hbox{Im} ~ #1}

\newcommand{\bea}{\begin{eqnarray}}
\newcommand{\eea}{\end{eqnarray}}
\newcommand{\ben}{\begin{enumerate}}
\newcommand{\een}{\end{enumerate}}
\newcommand{\bit}{\begin{itemize}}
\newcommand{\eit}{\end{itemize}}

\newcommand{\bert}{\raise-0.45mm\hbox{\Large$\Box$}}	%D'Alembertian

\definecolor{BrickRed}{cmyk}{0,0.89,0.94,0.28}					%%%PANTONE 1805
\definecolor{MidnightBlue}{cmyk}{0.98,0.13,0,0.43}				%%%PANTONE 302
\definecolor{DarkGreen}{rgb}{0.100806,0.495968,0.209979}
\definecolor{orange}{rgb}{0.587167,0.354498,0.146197}

\begin{document}

\title{Violation of Detailed Balance in Quantum Open Systems}
	
\author{Robert Alicki}
\email{robert.alicki@ug.edu.pl}
\affiliation{International Centre for Theory of Quantum Technologies (ICTQT), University of Gda\'{n}sk, 80-308, Gda\'{n}sk, Poland}
\author{Milan \v{S}indelka}
\email{sindelka@ipp.cas.cz}
\affiliation{Institute of Plasma Physics of the Czech Academy of Sciences, Za Slovankou 1782/3, 18200 Prague, Czech Republic}
\author{David Gelbwaser-Klimovsky}
\email{dgelbi@technion.ac.il}
\affiliation{Schulich Faculty of Chemistry and Helen Diller Quantum Center, Technion-Israel Institute of Technology, Haifa 3200003, Israel}

\date{\today}

\begin{abstract}
We consider the dynamics of a quantum system immersed in a dilute gas at thermodynamic
equilibrium using a quantum Markovian master equation derived by applying the low-density
limit technique. It is shown that the Gibbs state at the bath temperature is always stationary while
the detailed balance condition at this state can be violated beyond the Born approximation. This
violation is generically related to the absence of time-reversal symmetry for the scattering T matrix,
which produces a  thermalization mechanism that allows the presence of persistent probability and
heat currents at thermal equilibrium. This phenomenon is illustrated by a model of an electron
hopping between three quantum dots in an external magnetic field.

\end{abstract}


\maketitle

%%%%%%%%%%
%%% INTRODUCTION
%%%%%%%%%%
\section{Introduction}
Detailed balance at equilibrium (DBE) \cite{lewis_new_1925,fowler_note_1925,spohn_entropy_1978} is  a core principle of today´s thermodynamics. It ensures the lack of persistent currents at equilibrium \cite{zia_probability_2007} and plays a key role in a wide range of ﬁelds, including the Onsager relations \cite{onsager_reciprocal_1931,onsager_reciprocal_1931-1} and reaction kinetics \cite{boyd_macroscopic_1977} in chemistry, ﬂuctuation theorems \cite{kubo_fluctuation-dissipation_1966,cengio_fluctuation-dissipation_2021,crooks_entropy_1999,crooks_nonequilibrium_1998,kurchan_fluctuation_1998} in statistical mechanics, open quantum systems \cite{kossakowski_quantum_1977,alicki_detailed_1976-1} in quantum mechanics, and  Kirchhoﬀ’s law \cite{snyder_thermodynamic_1998} in electromagnetism. Detailed balance has been so closely identiﬁed with thermal equilibrium \cite{dann_open_2021} that its violation has been used as an indicator of lack of equilibrium \cite{battle_broken_2016,gnesotto_broken_2018} and has also been  suggested as a measure of distance from equilibrium \cite{platini_measure_2011}.

The assumption of DBE is prevalent across several ﬁelds. But interestingly, it is not actually required by any fundamental law \cite{thomsen_logical_1953}. This was long ago recognized by Onsager \cite{onsager_reciprocal_1931} himself who brought up the Hall eﬀect as an example where this principle does not hold. Other examples of systems that violate detailed balance include the Michaelis-Menten kinetics for enzyme kinetics \cite{johnson_original_2011,voorsluijs_thermodynamic_2020}, totally asymmetric simple exclusion process for one-dimensional transport \cite{macdonald_kinetics_1968,derrida_exact_1993}  directed percolation for fluid dynamics \cite{hinrichsen_non-equilibrium_2000} and nonreciprocal systems \cite{caloz_electromagnetic_2018,asadchy_tutorial_2020,zhu_near-complete_2014}. Many unexpected   effects in nonreciprocal materials have been theoretically predicted in the last  years: persistent heat currents in thermal equilibrium \cite{zhu_persistent_2016}, violations of the Kirchhoﬀ law \cite{zhu_near-complete_2014}, potential violations of  Earnshaw's theorem \cite{gelbwaser-klimovsky_equilibrium_2022}, deviations from the Green-Kubo relations \cite{herz_green-kubo_2019}, photon thermal Hall eﬀect \cite{ben-abdallah_photon_2016},  giant magnetoresistance for the heat ﬂux \cite{latella_giant_2017}, and the creation of a Casimir heat engine \cite{gelbwaser-klimovsky_near_2021}.

Unfortunately, currently used tools are insufficient for developing the microscopic models needed to study the dynamics and thermodynamics of systems that violate DBE. Here, we use the following definition of DBE
 \begin{gather}
 a_{k\ell}e^{-\beta \mathcal{E}_{\ell}} = a_{\ell k} e^{-\beta\mathcal{E}_k},
 \label{minDB}
 \end{gather}
where  $a_{k\ell}$ are transition rates between microstates  of the system with energies $\mathcal{E}_k , \mathcal{E}_{\ell}$ and $\beta$ is the inverse temperature of the bath. The Gorini, Kossakowski, Lindblad and Sudarshan (GKLS) \cite{lindblad_generators_1976,  gorini_completely_1976} equation derived at the weak coupling limit \cite{davies_markovian_1974} cannot be used because it automatically complies with DBE (see Ref. \cite{alicki_quantum_2007-1} and section III). The lack of microscopic models has resulted in contradicting statements in the literature regarding basic thermodynamic properties, such as the possibility of reaching thermal equilibrium \cite{asadchy_tutorial_2020,reimann_introduction_2002,hanggi_stochastic_1982,mann_nonreciprocal_2020,khandekar_new_2020} and the divergence of entropy production \cite{ben-avraham_entropy_2011-1,zeraati_entropy_2012,saha_entropy_2016,murashita_nonequilibrium_2014,raz_mimicking_2016}. 

To clarify the thermodynamic properties of systems that violate DBE, we use a GKLS master equation  in the low density limit (LDL) \cite{dumcke_low_1985}  which can lead into the violation of DBE. We note that the notion of DBE is not restricted to this limit and it will be interesting to study its violation beyond this regime. We prove that despite the violation of DBE,  fundamental thermodynamic behavior still holds: the reduced system reaches thermal equilibrium, at which the  entropy production is zero. Nevertheless,   DBE violation produces a different thermalization mechanism that allows persistent probability and heat currents at thermal equilibrium. To exemplify these effects we study a toy model for a single electron tunneling between  three quantum dots  in the presence of a magnetic field.

\section{Quantum Master Equations at low density limit}

We consider a   quantum system $S$ with a discrete spectrum physical Hamiltonian, 
$
H_S = \sum_k \mathcal{E}_k |k\rangle\langle k|,
\label{HamS}
$
immersed in an ideal bosonic or fermionic gas of  free  (quasi-) particles at a thermal equilibrium state given by the inverse temperature $\beta$ and particle density $\nu$. The derivation of the dynamics for the reduced density matrix is performed under the assumption that the density of gas particles is low. As shown in Ref. \cite{dumcke_low_1985} this assumption  implies that the form of the master equation does not depend on particle statistics and is fully determined by  the scattering of a single particle by the system $S$. Therefore, it is sufficient to determine the Hamiltonian $H_{tot}$ of the system composed  of $S$ and a single particle, 
$
H_{tot} = H_S + H_P + H_{int}.
\label{Ham1}
$
We assume for simplicity that the gas particle is spinless and is described by  its Hamiltonian $H_P$ in momentum representation, 
\begin{equation}
H_P = \int d\mathbf{p}\, E_{\mathbf{p}} |\mathbf{p}\rangle\langle\mathbf{p}|, \quad \langle \mathbf{p}|\mathbf{p}'\rangle = \delta(\mathbf{p} -\mathbf{p}`).
\label{HamP}
\end{equation}
The state of the gas is described by the single-particle probability distribution in momentum space $G(\mathbf{p})$.

The  single-particle scattering   M\o ller wave operator is defined as \cite{taylor_scattering_2006} 
\begin{equation}
\Omega_+ =  \lim_{t\to\infty} e^{-iH_{tot} t} e^{i(H_S +H_P)t},
\label{Moller}
\end{equation}
and its superoperator version is  $\Gamma_+ = \Omega_+ \cdot \Omega_+^{\dagger} $. The  $T$ operator  is the main mathematical object describing the scattering process and   is  defined as
\begin{equation}
T = H_{int}\Omega_+. 
\label{Tmatrix}
\end{equation}
It produces a family of transition operators acting on the Hilbert space of $S$ and labeled by the Bohr frequencies of $H_S$ denoted by $\{\omega\}$ and pairs of the particle's momenta,
\begin{equation}
T_{\omega}(\mathbf{p}' , \mathbf{p}) =\sum _{\mathcal{E}_k - \mathcal{E}_{\ell} = \omega} \langle k , \mathbf{p}'| T |\mathbf{p}, \ell\rangle |k\rangle\langle\ell |.
\label{Tomega}
\end{equation}
We further assume that the dilute ideal gas is at a stationary state fully characterized  by the probability distribution in momentum space $G(\mathbf{p})$ and the particle density $\nu$. As proven in Ref. \cite{dumcke_low_1985} the reduced  dynamics of $S$  is governed by the following quantum master equation (QME)
\begin{equation}
\frac{d}{dt}\rho_S = -i [H_S , \rho_S ] + \mathcal{L}\rho_S,
\label{LDL_QME}
\end{equation}
where the dissipative generator is
\begin{gather}
 \mathcal{L}\rho_S = \nu \pi\sum_{\omega} \int d\mathbf{p}\int d\mathbf{p}'\, G(\mathbf{p}) \delta (E_{\mathbf{p}'} - E_{\mathbf{p}} +\omega ) \times \notag \\
 \bigl\{[ T_{\omega}(\mathbf{p}' , \mathbf{p})\rho_S , 
T^{\dagger}_{\omega}(\mathbf{p}' , \mathbf{p})] +[ T_{\omega}(\mathbf{p}' , \mathbf{p}) ,\rho_S 
T^{\dagger}_{\omega}(\mathbf{p}' , \mathbf{p})]\bigr\} .
\label{LDL_gen}
\end{gather}
$\mathcal{L}$ can be expressed in the form of an ergodic average 
\begin{equation}
	\mathcal{L}= \lim_{a\to\infty} \frac{1}{a} \int_0^a dt e^{i t[H_S, \cdot]} \mathcal{L}_0 \,  e^{-i t[H_S, \cdot]}, 
	\label{averaging}
\end{equation}
where $\mathcal{L}_0$ is given by
\begin{gather}
	\mathcal{L}_0 \rho_S =-i \int d\mathbf{p}\, \langle\mathbf{p}| [H_{int}, \Gamma_+ (\rho_S\otimes\rho_P)]|\mathbf{p}\rangle.
	\label{pregen}
\end{gather}
 $\rho_P$ is the formal density matrix for the gas particle. 
This averaging (Eq. \eqref{averaging}) is usually associated with the \emph{secular approximation} which is a necessary step to assure positivity preserving of the derived QME.

The basic properties of the QME given by Eqs. \eqref{LDL_QME} and \eqref{LDL_gen} are the following.

(1) The dissipative generator  $\mathcal{L}$ commutes with the Hamiltonian part $-i[H_S, \cdot]$,
\begin{equation}
\mathcal{L}[H_S, \cdot] = [H_S, \cdot]\mathcal{L}. 
\label{commutation}
\end{equation} 
This  implies that populations of $H_S$ eigenstates evolve independently of their coherences.

(2) If the gas is at thermal equilibrium at the inverse temperature $\beta$ the probability distribution  of the particle's momenta is given by
\begin{equation}
G(\mathbf{p}) = Z^{-1} e^{-\beta E_{\mathbf{p}}}
\label{thermaleq}
\end{equation}
and the stationary state of the system  is  the Gibbs state, 
\begin{equation}
\rho^{\beta}_{S} = Z_S^{-1} e^{-\beta H_S}.
\label{Gibbs}
\end{equation}
\par

(3) Under the additional \emph{ergodicity condition}, any initial state of the system relaxes to the Gibbs state $
\rho_S^\beta$.

\textbf{Proofs}\\
Property (1) is a direct consequence of the averaging procedure [Eq. \eqref{averaging}]. Namely, using the following identity, valid for any fixed $\tau$
\begin{gather}
	\mathcal{L}= \lim_{a\to\infty} \frac{1}{a} \int_0^a dt e^{i (t+\tau)[H_S, \cdot]} \mathcal{L}_0 \,  e^{-i (t+\tau)[H_S, \cdot]} = \notag\\
	e^{i \tau[H_S, \cdot]}\mathcal{L}e^{-i\tau[H_S, \cdot]}
	\label{averaging2}
\end{gather}
and differentiating both sides of Eq. \eqref{averaging2} at $\tau = 0$ one obtains Eq. \eqref{commutation}.
\par
Property (2) is a new result, as in Ref. \cite{dumcke_low_1985}  it is assumed that the system complies with microreversibility. This implies DBE.  Here, we use only the intertwining property of the wave operator $\Omega_+$,
\begin{equation}
\Omega_+ (H_S + H_P) \Omega^{\dagger}_+ =  H_S + H_P + H_{int} = H_{tot},
\label{Omegaprop}
\end{equation} 
or equivalently,
\begin{equation}
\Omega_+ e^{-\beta(H_S + H_P)} \Omega^{\dagger}_+ =  e^{-\beta H_{tot}}.
\label{Omegaprop1}
\end{equation} 
 $\mathcal{L} \rho_S^{\beta} = 0$ is obtained by assuming the gas particle is in a thermal state, $\rho_P= Z^{-1} e^{-\beta H_P}$  and using Eqs.\eqref{averaging}, \eqref{Omegaprop1} and \eqref{pregen} (below $C$ is an irrelevant constant). $\mathcal{L}  \rho_S^{\beta}$ is equal to
\begin{gather}
 C\lim_{a\to\infty}\frac{1}{a}\int_0^a dt\, e^{iH_S t}\mathrm{Tr}_P [H_{int}, \Omega_+ e^{-\beta(H_S + H_P)}\Omega^{\dagger}_+]e^{-iH_S t}
\end{gather}
Using Eq., \eqref{Omegaprop1}, $H_{int}=H_{tot}-H_S-H_P$, and the fact that $[H_{tot},e^{-\beta H_{tot}}]=0$ we get 
\begin{gather}
C \lim_{a\to\infty}\frac{ -1}{a}\int_0^a dt\, e^{iH_S t}\mathrm{Tr}_P [H_S + H_P,  e^{-\beta H_{tot}}] e^{-iH_S t} =  \notag \\
 C \lim_{a\to\infty}\frac{ -1}{a}\int_0^a dt\, [H_S , e^{iH_S t}\mathrm{Tr}_P ( e^{-\beta H_{tot}} )e^{-iH_S t}]. \label{eq:firstproof}
\end{gather} 
In the last equality we have used that the trace of a commutator is zero.   Equation \eqref{eq:firstproof} can be rewritten as
\begin{gather}
 C\lim_{a\to\infty}\frac{ -1}{a}\int_0^a dt\, \frac{d}{dt}\Bigl( e^{iH_S t}\mathrm{Tr}_P ( e^{-\beta H_{tot}} ) e^{-iH_S t}\Bigr) = C\times \notag\\
 \lim_{a\to\infty}\frac{ -1}{a}\Bigl\{ e^{iH_S a}\mathrm{Tr}_P ( e^{-\beta H_{tot}}) e^{-iH_S a} -\mathrm{Tr}_P  e^{-\beta H_{tot}} \Bigr\}
 = 0, \label{eq:thproof}
\end{gather}
where we use the fact that the numerator has finite norm.

Property (3) is a consequence of the results obtained in Ref. \cite{frigerio_stationary_1978}.
\par

Properties  (1)-(3) show that  the Gibbs state is the steady state of the QME obtained in LDL for thermal equilibrium environments (ideal gas) without any additional assumptions such as DBE or microreversibility.

\section{Detailed balance condition for LDL dynamics}

In this section we discuss the  sufficient generic conditions leading to the detailed balance condition  [Eq. \eqref{minDB}] for QME of the LDL type  [Eqs. \eqref{LDL_QME} and\eqref{LDL_gen}]. The analysis is much simpler for the case of an $H_S$ with nondegenerated spectrum.
\par
For an $H_S$ with a nondegenerated spectrum the diagonal elements  of the density matrix,  $p_k \equiv\langle k|\rho_S|k\rangle$, evolve independently from the off-diagonal ones and satisfy the Pauli master equation of the form
\begin{equation}
\frac{d}{dt} p_k = \sum_{\ell}\bigl( a_{k\ell}\, p_{\ell} -  a_{\ell k}\, p_k\bigr)
\label{PauliME}
\end{equation}
with
\begin{gather}
a_{k\ell} = \nu \pi \int d\mathbf{p}\int d\mathbf{p}'\, G(\mathbf{p}) \delta \bigl\{(E_{\mathbf{p}'} +\mathcal{E}_k)- ( E_{\mathbf{p}} +\mathcal{E}_{\ell} )\bigr\} \times \notag \\ |\langle k , \mathbf{p}'| T |\mathbf{p}, \ell\rangle|^2 .
\label{transprob}
\end{gather}

Using Eq. \eqref{transprob} with $G(\mathbf{p}) = Z^{-1} e^{-\beta E_{\mathbf{p}}}$ one derives the following  identity
\begin{equation}
a_{k\ell}e^{-\beta \mathcal{E}_{\ell}} = a_{\ell k} e^{-\beta\mathcal{E}_k} I(k,\ell),
\label{transprobsym}
\end{equation}
where
\begin{gather}
I(k,\ell)= \notag \\
\frac{\int d\mathbf{p}\int d\mathbf{p}'\, e^{-\beta E_{\mathbf{p}}} \delta \bigl\{E_{\mathbf{p}'} +\omega_{kl}-  E_{\mathbf{p}} \bigr\} |\langle k , \mathbf{p}'| T |\mathbf{p}, \ell\rangle|^2}{\int d\mathbf{p}\int d\mathbf{p}'\, e^{-\beta E_{\mathbf{p}}} \delta \bigl\{E_{\mathbf{p}'} +\omega_{kl}-  E_{\mathbf{p}} \bigr\} |\langle \ell , \mathbf{p}| T|\mathbf{p}', k\rangle|^2} .
\label{DBcond}
\end{gather}
Here $\omega_{kl}=\mathcal{E}_k-\mathcal{E}_{\ell}$.
The DBE condition is satisfied if and only if $I(k, \ell) = 1$ for those pairs $(k,\ell)$ for which transition probabilities are nonzero. It may happen incidentally for a particular choice of the parameters, but we discuss only the generic situations which are related to  symmetries of the system.

The first  sufficient symmetry condition is  \emph{Hermicity} of the $T$ matrix $(T = T^{\dagger})$, that is $\langle k , \mathbf{p}'| T |\mathbf{p}, \ell\rangle =\overline{ \langle \ell , \mathbf{p}| T|\mathbf{p}', k\rangle}$. This is always satisfied for the Born approximation where  $ T \simeq H_{int}$. This approximation is valid at the weak coupling limit where DBE always holds. Physically, at the dilute limit, a hermitian  $T$ matrix represents a lossless system \cite{caloz_electromagnetic_2018}. 
\par
The second sufficient condition is assuming that the T matrix is a symmetric matrix, $\langle k , \mathbf{p}'| T|\mathbf{p}, \ell\rangle=\langle \ell , \mathbf{p}| T|\mathbf{p}', k\rangle$, which implies that the system is reciprocal \cite{asadchy_tutorial_2020,caloz_electromagnetic_2018}.
\par
The third  case corresponds to  \emph{time-reversal symmetry}  or microreversibility. It means that the states $|k\rangle$ are invariant with respect to time reversal, $E_{\mathbf{p}}=E_{-\mathbf{p}}$ and the probability of the scattering event 
$|\ell , \mathbf{p}\rangle\mapsto |k , \mathbf{p}'\rangle$ is equal to the probability of time-reversed event  $|k , -\mathbf{p}'\rangle \mapsto |\ell , -\mathbf{p}\rangle$. This condition means
\begin{equation}
|\langle k , \mathbf{p}'| T |\mathbf{p}, \ell\rangle|^2=  |\langle \ell , -\mathbf{p}| T|-\mathbf{p}', k\rangle|^2 ,
\label{timereversal}
\end{equation}
which leads to $I(k,\ell) = 1$.
\par
The fourth condition combines time reversal with parity transformation (space inversion) which leads to the condition
\begin{equation}
|\langle k , \mathbf{p}'| T |\mathbf{p}, \ell\rangle|^2=  |\langle \ell , \mathbf{p}| T|\mathbf{p}', k\rangle|^2.
\label{parity}
\end{equation}
We note here that only  on shell processes have to be considered. This is a consequence of the delta function in Eq. \eqref{transprob}, which ensures energy conservation. For particular systems it may happen that certain  geometric symmetry can restore detailed balance (see a toy model in Section V and Supplementary Information (SI).

The above conditions were derived for the T matrix. Some of them can also be obtained for the Green's operator (see discussion below Eq S9 on the SI).

Fulfilling at least one of the mentioned conditions, will be enough to ensure DBE. In section V we show a toy model that does not comply with any of the   above conditions, resulting in DBE violation (see Section S.2.a on SI).

\section{Thermodynamic laws and entropy production}

DBE violation  provides additional freedom to the reduced dynamics. Nevertheless,  the time invariance of the Gibbs state still allows for preserving the fundamental principles of thermodynamics: the impossibility of steady work extraction from a single thermal bath or  cooling of a cold bath without an external driving. Mathematically, the LDL  master equation   \eqref{LDL_QME}, satisfies:  (i) the zeroth law of thermodynamics  [see Eq. \eqref {thermaleq}], (ii)  the  first law of thermodynamics (implied by the Hamiltonian model of open system), and (iii) the second law of thermodynamics (implied by  the Spohn inequality \cite{spohn_entropy_1978}).
\par
For diagonal density matrices the entropy production defined as
\begin{gather}
\sigma =    \sum_{k}\frac{dp_k}{dt}[\ln p^{eq}_k - \ln p_k]; \quad    p^{eq}_k \equiv Z^{-1}e^{-\beta\mathcal{E}_k}, \label{eq:entprod}
\end{gather}
can be written in terms of the DBE violation as
\begin{gather}
    \sigma=\sum_{k>j}\left[K_{jk}
    \ln\left(\frac{p_k}{p_j}\frac{a_{jk}}{a_{kj}}\right)+K_{jk}\ln I(k,j)\right]. \label{eq:entr}
\end{gather}
where $K_{jk}=a_{jk}p_k -  a_{kj}p_j$ is the probability current from the microstate $\{k\}$ to the microstate  $\{j\}$.
The first term in Eq. \eqref{eq:entr} corresponds to the Schnakenberg formulation of entropy production \cite{schnakenberg_network_1976} and the second to deviations due to the lack of DBE. 
 Equation \eqref{eq:entr} is valid for  any temporal state as long as the  steady state of the dynamics is a Gibbs state [see Eq. \eqref{Gibbs}] as  is the case for Eq. \eqref{LDL_QME}.

 Furthermore, at thermal equilibrium the second term can be rewritten as 
\begin{gather}
    \sum_{k>j}a_{jk}p^{eq}_k\left[1-I(k,j)\right]\ln I(k,j), \label{eq:devterm}
\end{gather}
which is always negative.   At thermal equilibrium the first and second laws force the entropy production to be zero. Therefore, to compensate for the negativity of Eq. \eqref{eq:devterm},
the  Schnakenberg entropy production should be positive and increases with the DBE violation:  $|1-I(k,j)|$.

\section{Three-level open system without time-reversal symmetry}

As a toy model we consider a single (spinless) electron that can occupy  three dots ($1$,$2$,$3$)  in an equilateral triangle arrangement in the presence of a magnetic field. 
The positions of the quantum dots (QD) are given by $\{\mathbf{q}_j ; \iota =1,2,3\}$. This system  (3QD) is governed by the single particle Hubbard Hamiltonian. This Hamiltonian type has been used to study more complex systems, such as Benzene molecules \cite{schuler2013optimal} and could be used to extend our results to more realistic scenarios. In the single electron localized basis ($|1\rangle, |2\rangle, |3\rangle$), the Hamiltonian is given by
\begin{equation}
H_{el}=\tau\left(\begin{array}{ccc}
0 & e^{-i2\pi\phi/3} & e^{i2\pi\phi/3}\\
e^{i2\pi\phi/3} & 0 & e^{-i2\pi\phi/3}\\
e^{-i2\pi\phi/3} & e^{i2\pi\phi/3} & 0
\end{array}\right),    
\label{eq:hamel}
\end{equation}
where $\tau$ is the tunneling parameter and $\phi$ is the magnetic flux \cite{delgado_theory_2007}. 
The diagonal form of this Hamiltonian is $H_{el}=\sum_{j\in \{\pm,0\}}=\mathcal{E}_j |j\rangle \langle j|$ where $\mathcal{E}_j=-2|\tau|\cos[2\pi (\phi+j*1)/3]$. For most values of the magnetic flux the Hamiltonian is nondegenerated.

\begin{figure}
    \centering
    \includegraphics[width=0.35\textwidth, angle=-90]{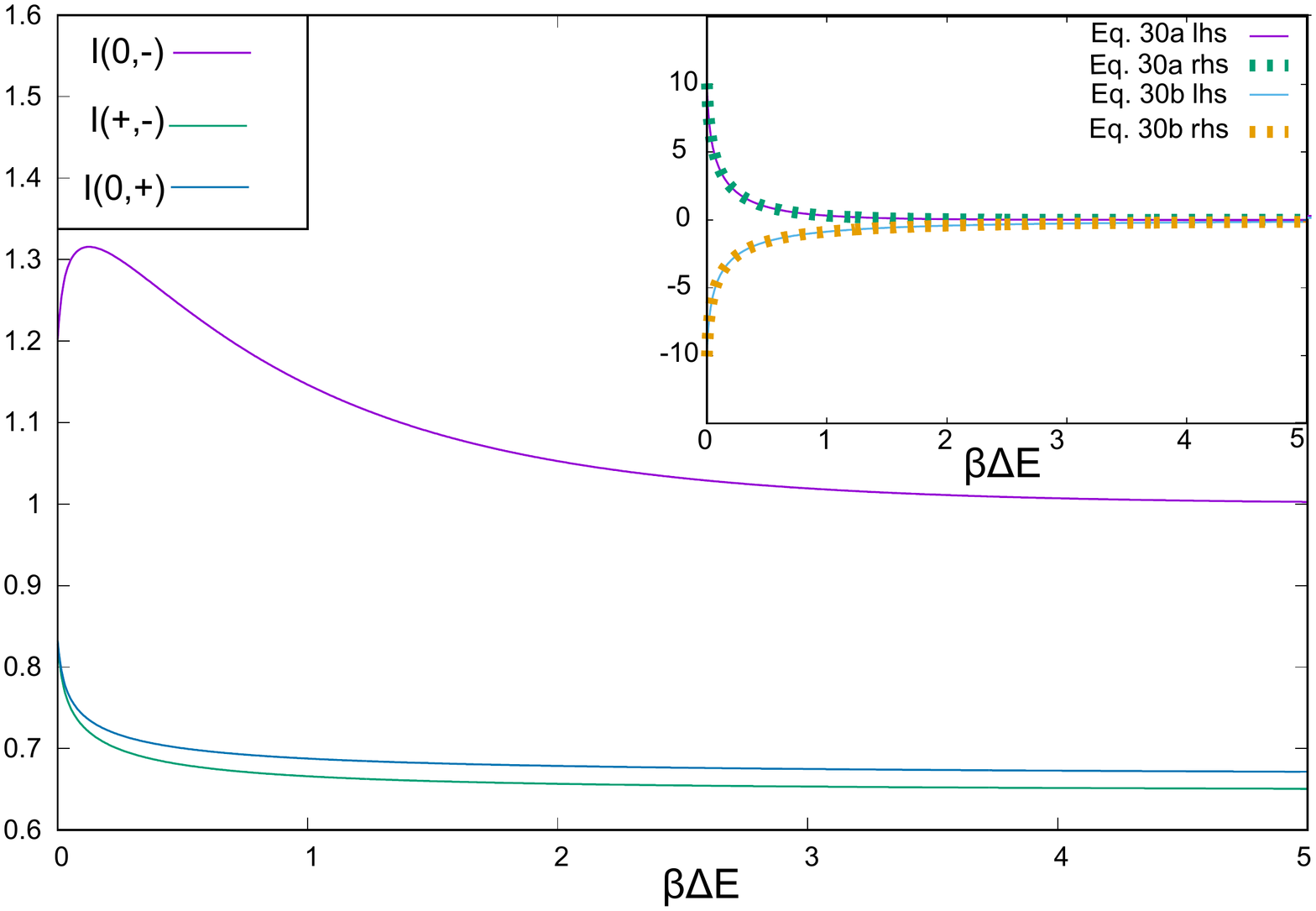}
    \caption{$I(k,l)$ as function of the normalized inverse temperature $\beta \Delta E$. DBE only holds if all the $I(k,l)=1$. For large $\beta \Delta E$, $I(0,-)$ tends to 1. Nevertheless,  $I(+,-)$ and $I(0,+)$ are still different from 1 indicating the lack of DBE. Inset: Left hand side, lhs, (continuous line) and  right hand side, rhs (dotted line), of the thermalization conditions, Eqs. 30a and 30b. The compliance with these conditions (lhs=rhs) verifies our numeric calculation. The y-axis of the inset has been multiplied by a factor of $10^4$. Parameters: ${\cal V}_1=1$,${\cal V}_2=0.7$, ${\cal V}_3=1.5$, $\mathcal{E}_+=0.5$, $\mathcal{E}_0=0.0$,  $\mathcal{E}_-=-0.5$ and $\Delta E=\mathcal{E}_0-\mathcal{E}_-$. For more details see SI.}
    \label{fig:num}
\end{figure}

The 3QD interacts with a low-density gas of free particles of mass $m$ that is at a thermal state with  inverse temperature $\beta$. We assume a simple form of the interaction potential given by 
\begin{equation}
H_{int}(\mathbf{q})= \sum_{\iota\in \{1,2,3 \}}V_\iota(\mathbf{q}-\mathbf{q}_\iota)|\iota\rangle\langle \iota|,
\label{Intpotential}
\end{equation}
 where $V_\iota(\mathbf{q}-\mathbf{q}_\iota)$ is a short range repulsive potential between the electron and the particle. In order to simplify numerical calculations we assume that the distance between dots is small in comparison with the typical wavelength of a quantum scatterer. This allows one to treat all dots as sitting at the same point, while keeping the structure of the internal electron Hamiltonian [Eq. \eqref{eq:hamel}]. Then we can replace smooth potentials in Eq. \eqref{Intpotential} by 1D Dirac deltas ${\cal V}_\iota \delta(q)$ and at the same time use as a heat bath a one-dimensional particle gas. Figure \ref{fig:num} shows the results of numerical calculations for this simplified model showing a substantial violation of DBE while the stationary state remains a Gibbs one. Notice that in the presented example all coupling constants ${\cal V}_{\iota}$ are different. It is shown in the SI that if at least two constants are equal DBE is preserved. This is an example of a system-specific symmetry that restores DBE. Namely, here time reversal exchanges the eigenstates $|\pm\rangle$ that can be undone by the permutation of two states from the set $\{|\iota\rangle\}$ with equal couplings to the bath.

DBE establishes a relation between transition rates involving the same Bohr frequencies (i.e., $a_{kl}$ and $a_{lk}$)  allowing independent transition rates among different Bohr frequency \cite{dann_open_2021}. This relation, together with the Kubo-Martin-Schwinger condition (KMS), forces the reduced system to thermalize.   In contrast, systems violating DBE use a different thermalization mechanism. While they have some extra degree of freedom due to the DBE violation, thermalization imposes a complex dependence among rates for different Bohr frequencies that we term thermalization conditions. For example, in the case of the 3QD model they read
\begin{subequations}\label{thcond:main}
\begin{equation}
 a_{+0}\left[1-I(0,+)\right]=a_{-0}\left[I(0,-)-1\right];
 \label{thcond:a}
\end{equation}
\vspace{-1cm}
\begin{equation}
	a_{0+}\left[1-1/I(0,+)\right]=a_{-+}\left[I(+,-)-1\right].   
 \label{thcond:b}
\end{equation}
\end{subequations}

\subsection{Probabillity and heat currents}
The different thermalization mechanisms could be better understood by analyzing the probability currents, $K_{jk}$ (see discussion below Eq. \eqref{eq:entr} and Ref.\cite{zia_probability_2007}).
Systems complying with DBE thermalize by reducing each individual probability's currents, until all of them become zero at thermal equilibrium.
In contrast, systems violating DBE thermalize by reducing $\sum _j K_{jk}$ which becomes zero at thermal equilibrium, while at least some of the individual currents, 
 \begin{equation}
	K_{jk} =  a_{kj}p^{eq}_k \left[I(j,k)-1\right],
	\label{probcurrent}
\end{equation}
 remain nonzero even at equilibrium forming closed loops. These persistent currents are different from those found on aromatic \cite{merino2004induced,gomes2001aromaticity,johansson2005sphere} or mesoscopic rings \cite{von1991average,chen_nonequilibrium_2016}. The currents found in these works are also present in isolated systems. They  are  produced by breaking the time-reversal symmetry of the system eigenfunctions \cite{shanks2011persistent}. In contrast, the current described by Eq. \eqref{probcurrent} requires   a nonisolated system and the  breakdown of other symmetries (see section III).

It has been claimed that  violation of DBE produces persistent heat currents  in nonreciprocal systems \cite{zhu_persistent_2016}.
The existence of these currents  does not violate any fundamental thermodynamic law. Using the Spohn inequality it is possible to define  thermodynamically consistent heat currents \cite{gelbwaser2015} between individual pairs of the system energy levels  and the thermal bath. At  the equilibrium state they are defined as $J_{ml}=-\beta^{-1}{a_{lm}p^{eq}_m}\left[I(m,l)-1\right]\ln \left( p_m^{eq}/p_l^{eq}\right)$. In particular for the toy model we get ($i,j = 0, +, -$)
\begin{gather}
    J_{ij}=N \left[I(+,-)-1\right](\mathcal{E}_i-\mathcal{E}_j),    
\end{gather}
where $N=a_{-+}e^{-\beta (\mathcal{E}_--\mathcal{E}_0)}[1+e^{-\beta (\mathcal{E}_--\mathcal{E}_0)}+e^{-\beta (\mathcal{E}_+-\mathcal{E}_0)}]^{-1}$. Even though there are heat currents to and from individual pairs of  the system energy levels, the total heat exchange between the bath and the system is zero, as expected from the first law of thermodynamics. It is important to notice that both probabilities and heat currents are related to transitions between delocalized energy levels and do not necessarily imply the existence of heat  currents in space.  In particular, for our 3QD model all energy eigenstates yield equal occupation probabilities for all dots.

In summary, here we develop an open quantum system framework to study systems that violate DBE. This extra degree of freedom changes the system dynamics and could be beneficial for many applications, such as: speeding up thermalization, increasing the sensitivity of measuring devices, and improving the operation of heat machines. One should stress that DBE is not easy to break. The effect appears in the higher-order expansion with respect to the system-bath coupling constant and  could vanish in the presence of certain spatial symmetries (see SI).

\textit{Acknowledgement}
We thank Natan Granit, Thales Pinto Silva, U. Peskin and N. Moiseyev  for useful discussions. R.~A.~is supported by the Foundation for Polish Science's International Research Agendas, with structural funds from the European Union (EU) for the ICTQT and at the Technion by a fellowship from the Lady Davis Foundation. M.~\v{S}.~gratefully acknowledges financial support of the Helen Diller Quantum Center of the Technion (March 1 - May 31, 2023).  D.G.K. is  supported by the ISRAEL SCIENCE FOUNDATION (grant No. 2247/22) and by the Council for Higher Education Support Program for Hiring Outstanding Faculty Members in Quantum Science and Technology in Research Universities.

%\bibliography{zotero2}

\setcounter{equation}{0}
\renewcommand{\theequation}{S\arabic{equation}}

\setcounter{figure}{0}
\renewcommand{\thefigure}{S\arabic{figure}}

\setcounter{section}{0}
\renewcommand{\thesection}{S\arabic{section}}

\setcounter{subsection}{0}
\renewcommand{\thesubsection}{\thesection.\alph{subsection}}

\onecolumngrid
\vspace{1cm}
\begin{centering}
{\large \bf Violation of Detailed Balance in Quantum Open Systems: Supplementary information}\\

\end{centering}

\vspace{1cm}
The purpose of this supplementary information is to introduce a simple toy model which is used in the main text to explicitly illustrate the violation of detailed balance. We start with a rather general setup and go subsequently into a more specific setting which enables us to obtain $I(k,l)$ semianalytically without any approximations.

\section{General setup for the toy model}

%---------------------------------------------------------------------------------------------------------------------------

 We consider a non-relativistic  quantum particle  in a $d$-dimensional space.
      The corresponding Hilbert space is spanned  by the momentum basis  $| {\bm p} \ra$, or equivalently by the position basis $| {\bm q} \ra$. For a joined system-particle basis one has
      \be
         \la {\bm q} \,\n j | {\bm p} \,\n j'\n \ra \; = \; \delta_{jj'} \, \frac{e^{\frac{i}{\hbar}{\bm p}\bm\cdot{\bm q}}}{(2\,\pi\,\hbar)^{d/2}}  .
      \ee
      State vectors $|\Psi\ra$ can be represented by wavefunctions
      \be
         \Psi_j({\bm q}) \; = \; \la {\bm q} \,\n j | \Psi \ra , \mez \tilde{\Psi}_j({\bm p}) \; = \; \la {\bm p} \,\n j | \Psi \ra  .
      \ee
      The Hamiltonian $H_0=H_S+H_P$ of our two isolated subsystems satisfies
      \be
         H_0 \, | {\bm p} \,\n j \ra \; = \; \Bigl( E_p + {\cal E}_j \Bigr) \, | {\bm p} \,\n j \ra ;
      \ee
      here $E_p=\frac{p^2}{2\,m}$. The interaction among the two subsystems,  $H_{int}$ is defined by prescription
      \be \label{coupling}
         \la {\bm p} \,\n j |H_{int} | \Psi \ra \; = \; \Theta(P-p) \, \sum_{j'} \sum_\iota \, v_{jj'}^{\iota,P} \, \Psi_{j'}\n({\bm q}_\iota)
         \; \frac{e^{-\frac{i}{\hbar}{\bm p}\bm\cdot{\bm q}_\iota}}{(2\,\pi\,\hbar)^{d/2}};
      \ee
      here $P$ is the momentum UV cutoff needed for eventual renormalization\footnote{
      Renormalization is not needed for $d=1$, but it is inevitable for $d=2$ and $3$, see \cite{jackiw1991delta}. }.
      One may ask at this point where the particular form (\ref{coupling}) of the coupling comes from.\\
      Here is the answer:
      \begin{eqnarray}
         & & \la {\bm p} \,\n j | \left( \sum_\iota \, | \chi_\iota \ra \, {\cal V}_\iota \; \delta^{d}\n(\hat{\bm q}-{\bm q}_\iota) \; \la \chi_\iota | \right)
         \int_{{\mathbb R}^d} \m {\rm d}^d\n q' \sum_{j'} \Psi_{j'\n}({\bm q}'\n) \; | {\bm q}'\n \,\n j'\n \ra \; = \nonumber\\
         & = & \sum_{j'} \sum_\iota \, \Bigl( \,\m _{\rm }\la j | \chi_\iota \ra \, {\cal V}_\iota \,\la \chi_\iota | j'\n \ra_{\rm } \Bigr) \,
         _{\rm el}\la {\bm p} | {\bm q}_\iota \ra_{\rm el} \, \Psi_{j'}\n({\bm q}_\iota).\label{eq:transfer}
      \end{eqnarray}

 Let us look now at the implicit Lippmann-Schwinger equation (LSE)

         \be \label{LSE-implicit}
         | \Psi \ra \; = \; | {\bm p} \,\n j \ra \; + \; G_0 \, H_{int} \, | \Psi \ra.
      \ee
      Here $G_0$ is the free Green's operator, $G_0=\frac{1}{E_p + {\cal E}_j -H_0 + i\varepsilon}$. The corresponding $T$-matrix elements are given simply as
      \be \label{T-matels-take-0}
         \la {\bm p}'\n \,\n j'\n | \,T \, | {\bm p} \,\n j \ra \; = \; \la {\bm p}'\n \,\n j'\n | \, H_{int} \, | \Psi \ra=\la {\bm p}'\n \,\n j'\n |H_{int}| {\bm p} \,\n j \ra +\la {\bm p}'\n \,\n j'\n |H_{int}G_0H_{int} | \Psi \ra.
      \ee
Moreover, we can rewrite the above two expressions in terms of the full Green's operator,  $G=\frac{1}{E_p + {\cal E}_j -H_{tot} + i\varepsilon}$. One has
     \be \label{LSE-explicit}
         | \Psi \ra \; = \; | {\bm p} \,\n j \ra \; + \; G\, H_{int} \, |  {\bm p} \,\n j \ra \; ;
      \ee
and
        \be \label{T-matels-take-0-explicit}
         \la {\bm p}'\n \,\n j'\n | \,T \, | {\bm p} \,\n j \ra \; =\la {\bm p}'\n \,\n j'\n |H_{int}| {\bm p} \,\n j \ra +\la {\bm p}'\n \,\n j'\n |H_{int}GH_{int} | {\bm p} \,\n j \ra \; .
      \ee
Equation (\ref{T-matels-take-0-explicit}) can be used to relate the $T-$matrix  properties with the full Green's operator and, in this way, determine the necessary conditions on the full Green's operator for DBE. For example, if $\la {\bm p}' j' | H_{int} | {\bm p} j \ra$ is a symmetric matrix, then the same applies also for the matrix elements of $G$, and therefore also the $T$-matrix elements are symmetric. Establishing the requirement on the Green's function for complying with the other conditions on the $T$-matrix (hermiticity, time reversal with and without parity, see section III on the main text) is left for future works.\\

    Let us return now to equation (\ref{T-matels-take-0}).
      The r.h.s.~can actually be evaluated explicitly as above in (\ref{coupling}).
      Meaning also that all the $T$-matrix elements are completely specified by knowledge of $\{ \Psi_j({\bm q}_\iota) \}_{j,\iota}$.
      Note also that
      \be \label{G-0-momentum-representation}
         \la {\bm p}'\n \,\n j'\n | \, \frac{1}{E_p + {\cal E}_j - H_0 + i\varepsilon} \, | {\bm p}''\n \,\n j''\n \ra \; = \;
         \frac{\delta^d\n({\bm p}'\n-{\bm p}''\n) \; \delta_{j'\n j''\n}}{E_p + {\cal E}_j - E_{p'\n} - {\cal E}_{j'\n} + i\varepsilon}.
      \ee
%---------------------------------------------------------------------------------------------------------------------------
 Returning to (\ref{LSE-implicit}) and taking advantage of (\ref{G-0-momentum-representation}) and (\ref{coupling}). One gets
      \begin{eqnarray} \label{LSE-implicit-take-2}
      %  -------------------------------------------------------------------------------------------------------------------
         \tilde{\Psi}_{j'\n}({\bm p}'\n) & = & \delta^d\n({\bm p}-{\bm p}'\n) \; \delta_{jj'} \; + \;
         \int_{{\mathbb R}^d} \m {\rm d}^d\n p'' \sum_{j''} \, \frac{\delta^d\n({\bm p}'\n-{\bm p}''\n) \;
         \delta_{j'\n j''\n}}{E_p + {\cal E}_j - E_{p'\n} - {\cal E}_{j'\n} + i\varepsilon} \;
         \la {\bm p}''\n \,\n j''\n | H_{int} | \Psi \ra \; = \nonumber\\
      %  -------------------------------------------------------------------------------------------------------------------
         & = & \delta^d\n({\bm p}-{\bm p}'\n) \; \delta_{jj'} \; + \; \frac{\la {\bm p}'\n \,\n j'\n | H_{int} | \Psi \ra}
         {E_p + {\cal E}_j - E_{p'\n} - {\cal E}_{j'\n} + i\varepsilon} \; = \nonumber\\
      %  -------------------------------------------------------------------------------------------------------------------
         & = & \delta^d\n({\bm p}-{\bm p}'\n) \; \delta_{jj'} \; + \; \frac{\Theta(P-p'\n)}{E_p + {\cal E}_j - E_{p'\n} - {\cal E}_{j'\n} + i\varepsilon}
         \; \sum_{j''} \sum_\iota \, v_{j'\n j''}^{\iota,P} \, \Psi_{j''}\n({\bm q}_\iota) \; \frac{e^{-\frac{i}{\hbar}{\bm p}'\n\bm\cdot{\bm q}_\iota}}
         {(2\,\pi\,\hbar)^{d/2}}.
      %  -------------------------------------------------------------------------------------------------------------------
      \end{eqnarray}
      Showing once again that the scattering wavefunction $\tilde{\Psi}_{j'\n}({\bm p}'\n)$ is known iff 
      a finite sequence of values of $\{ \Psi_{j''}\n({\bm q}_\iota) \}_{j''\n\iota}$ is known.
      In passing we note that there is no scattering for $p'\n>P$ due to presence of $\Theta(P-p'\n)$.
%---------------------------------------------------------------------------------------------------------------------------
 What remains to be done is to find the above mentioned values of $\{ \Psi_j({\bm q}_\iota) \}_{j,\iota}$.\\
      One has
      \be
         \Psi_{j'}\n({\bm q}_{\iota'\n}) \; = \; \int_{{\mathbb R}^d} \m {\rm d}^d\n p' \; \tilde{\Psi}_{j'\n}({\bm p}'\n) \;
         \frac{e^{+\frac{i}{\hbar}{\bm p}'\n\bm\cdot{\bm q}_{\iota'\n}}}{(2\,\pi\,\hbar)^{d/2}};
      \ee
      and (\ref{LSE-implicit-take-2}) provides immediately
      \begin{eqnarray} \label{LSE-implicit-take-3}
      %  -------------------------------------------------------------------------------------------------------------------
         \Psi_{j'}\n({\bm q}_{\iota'\n})  =  \frac{e^{+\frac{i}{\hbar}{\bm p}\bm\cdot{\bm q}_{\iota'\n}}}{(2\,\pi\,\hbar)^{d/2}} \; \delta_{jj'}
      %  -------------------------------------------------------------------------------------------------------------------
          +  \sum_{j''} \sum_{\iota''} \, \left\{ \, \frac{v_{j'\n j''}^{\iota''\m,P}}{(2\,\pi\,\hbar)^d} \int_{{\mathbb R}^d} \m {\rm d}^d\n p' \;
         \frac{\Theta(P-p'\n) \; e^{+\frac{i}{\hbar}{\bm p}'\n\bm\cdot({\bm q}_{\iota'\n}-{\bm q}_{\iota''\n})}}
         {E_p + {\cal E}_j - E_{p'\n} - {\cal E}_{j'\n} + i\varepsilon} \, \right\} \Psi_{j''}\n({\bm q}_{\iota''\n}).
      %  -------------------------------------------------------------------------------------------------------------------
      \end{eqnarray}
      The just obtained outcome (\ref{LSE-implicit-take-3}) represents
      a set of linear inhomogeneous equations for the unknowns $\{ \Psi_j({\bm q}_\iota) \}_{j,\iota}$.
      It can be solved either analytically or numerically, assuming tacitly regularity.
      Having the coefficients $\{ \Psi_j({\bm q}_\iota) \}_{j,\iota}$ in hand, all the $T$-matrix elements (\ref{T-matels-take-0}) can be accessed using (\ref{coupling}).
      Such that
      \be \label{T-matels}
         \la {\bm p}'\n \,\n j'\n | \, T \, | {\bm p} \,\n j \ra \; = \;
         \Theta(P-p'\n) \, \sum_{j''} \sum_{\iota''} \, v_{j'\n j''}^{\iota''\m,P} \, \Psi_{j''}\n({\bm q}_{\iota''}\n)
         \; \frac{e^{-\frac{i}{\hbar}{\bm p}'\n\bm\cdot{\bm q}_{\iota''}\n}}{(2\,\pi\,\hbar)^{d/2}}.
      \ee
      The final outcome of our above pursued analysis
      can be summarized by working equations (\ref{LSE-implicit-take-3}) and (\ref{T-matels}).
%---------------------------------------------------------------------------------------------------------------------------

\vspace*{+0.20cm}

\subsection{Short separation}

%---------------------------------------------------------------------------------------------------------------------------
Substantial simplification follows when all the sites ${\bm q}_\iota$ are placed at the origin ${\bm 0}$.\\
      One has
      \begin{eqnarray} \label{LSE-implicit-take-4}
      %  -------------------------------------------------------------------------------------------------------------------
         \Psi_{j'}  =  \frac{\delta_{jj'}}{(2\,\pi\,\hbar)^{d/2}} 
      %  -------------------------------------------------------------------------------------------------------------------
          +  \frac{1}{(2\,\pi\,\hbar)^d} \, \int_{{\mathbb R}^d} \m {\rm d}^d\n p' \;
         \frac{\Theta(P-p'\n)}{E_p + {\cal E}_j - E_{p'\n} - {\cal E}_{j'\n} + i\varepsilon}
         \; \sum_{j''} v_{j'\n j''}^{P} \; \Psi_{j''};
      %  -------------------------------------------------------------------------------------------------------------------
      \end{eqnarray}
      with
      \be
         v_{j'\n j''}^{P} \; = \; \sum_{\iota''} \, v_{j'\n j''}^{\iota''\m,P};
      \ee
      and $\Psi_{j'}\equiv\Psi_{j'}\n({\bm 0})$.
      Equation (\ref{T-matels}) for the $T$-matrix elements boils down into
      \be \label{T-matels-take-2}
         \la {\bm p}'\n \,\n j'\n | \, T \, | {\bm p} \,\n j \ra \; = \;
         \frac{\Theta(P-p'\n) }{(2\,\pi\,\hbar)^{d/2}} \, \sum_{j''} \, v_{j'\n j''}^{P} \, \Psi_{j''} .
      \ee
      Combining (\ref{T-matels-take-2}) and (\ref{LSE-implicit-take-4}) yields an even simpler formula
      \be \label{T-matels-take-3}
         \la {\bm p}'\n \,\n j'\n | \,T \, | {\bm p} \,\n j \ra \; = \; \Theta(P-p'\n) \,
         \left( \, \int_{{\mathbb R}^d} \m {\rm d}^d\n p''\n \; \frac{\Theta(P-p''\n)}{E_p + {\cal E}_j - E_{p''\n} - {\cal E}_{j'\n} + i\varepsilon} \, \right)^{\m\m\m-1}
         \Bigl( (2\,\pi\,\hbar)^{d/2} \, \Psi_{j'} \, - \, \delta_{jj'} \Bigr) .
      \ee
      The final outcome of the just presented analysis
      can be summarized by  equations (\ref{LSE-implicit-take-4}) and (\ref{T-matels-take-3}).
      Note that the $T$-matrix elements coming out of (\ref{T-matels-take-3}) are, by construction,
      independent upon the directions of ${\bm p}$ and ${\bm p}'$\m\footnote{
      This is not the case when formula (\ref{T-matels}) is used in the general setup. }.
  \\
%---------------------------------------------------------------------------------------------------------------------------

\subsection{ One dimensional case}

	%---------------------------------------------------------------------------------------------------------------------------
	 The UV cutoff $P$ can be lifted to $+\infty$, since for $d=1$ no renormalization is needed.
	Contour integration (residue theorem) provides the required integral
	\be \label{integral-1D}
	\int_{-\infty}^{+\infty} \m {\rm d}p'\n \; \frac{1}{E_p + {\cal E}_j - E_{p'\n} - {\cal E}_{j'\n} + i\varepsilon} \; = \;
	-\,i\,\pi\,\sqrt{\frac{2\,m}{E-{\cal E}_{j'}\n}};
	\ee
	here  $E = E_p + {\cal E}_j$.
	Note that $E-{\cal E}_{j'}\n>0$ as long as the $j'\n$-th channel is open for scattering,
	and $E-{\cal E}_{j'}\n<0$ as long as the $j'\n$-th channel is closed for scattering.
	For the sake of clarity, we present explicitly the calculation leading to (\ref{integral-1D}):
	\begin{eqnarray}
		%  -------------------------------------------------------------------------------------------------------------------
		\int_{-\infty}^{+\infty} \m \frac{{\rm d}p'}{E - {\cal E}_{j'\n} + i\varepsilon - E_{p'\n}} \; 
		 =  2 \, m \, \int_{-\infty}^{+\infty} \m \frac{{\rm d}p'}{2\,m\,(E - {\cal E}_{j'\n} + i\varepsilon) - p'^2} \; = 
		- \, 2 \, m \, \int_{-\infty}^{+\infty} \m \frac{{\rm d}p'}{(p'\n-y)(p'\n+y)}; 
		%  -------------------------------------------------------------------------------------------------------------------
	\end{eqnarray}
	where 
	\be
	y \; = \; \sqrt{2\,m\,(E - {\cal E}_{j'\n} + i\varepsilon)} , \mez \Re\,y > 0 , \mez \Im\,y > 0.
	\ee
	The integration contour over $p'$ can now be closed in the upper half of the complex $p'\n$-plane.
	In this way only the pole $p'\n=y$ is encircled, and the residue theorem yields accordingly
	\be
	\int_{-\infty}^{+\infty} \m \frac{{\rm d}p'}{E - {\cal E}_{j'\n} + i\varepsilon - E_{p'\n}} \; = \;
	-\,i\,\pi\,\sqrt{\frac{2\,m}{E-{\cal E}_{j'}\n+i\varepsilon}}.
	\ee
	Thereby (\ref{integral-1D}) is obtained for $\varepsilon \to +0$.
	%---------------------------------------------------------------------------------------------------------------------------
	 Equation (\ref{LSE-implicit-take-4}) boils down into
	\begin{eqnarray} \label{LSE-implicit-take-4-1D}
		%  -------------------------------------------------------------------------------------------------------------------
		\Psi_{j'} & = & \frac{\delta_{jj'}}{\sqrt{2\,\pi\,\hbar}} \; - \; \frac{i}{2\,\hbar} \, \sqrt{\frac{2\,m}{E-{\cal E}_{j'}\n}}
		\; \sum_{j''} v_{j'\n j''} \; \Psi_{j''}.
		%  -------------------------------------------------------------------------------------------------------------------
	\end{eqnarray}
	Equation (\ref{T-matels-take-3}) boils down into
	\be \label{T-matels-take-3-1D}
	\la p'\n \,\n j'\n | \,T \, | p \,\n j \ra \; = \; \frac{i}{\pi} \;
	\sqrt{\frac{E-{\cal E}_{j'}\n}{2\,m}} \, \Bigl( \sqrt{2\,\pi\,\hbar} \, \Psi_{j'} \, - \, \delta_{jj'} \Bigr).
	\ee

\section{Three levels system toy model}

In particular, we consider the three-level system described in the main text. Its Hamiltonian is defined in the main text Eq. 28. The localized basis, ($|1\rangle$, $|2\rangle$ and $|3\rangle$) corresponds to the states $|\chi_\iota\rangle$ in Eq. \ref{eq:transfer} of this supplement and $|j\rangle$, with $(j=-1,0,+1)$, to the three-level system eigenstates. Using the relation between the localized basis and the system Hamiltonian \cite{delgado_theory_2007}, one can find $v_{j'\n j''}$:
\begin{equation}
	v_{jj}=\frac{\mathcal{V}_{1}+\mathcal{V}_{2}+\mathcal{V}_{3}}{3};\label{eq:dia}
\end{equation}

\begin{equation}
	v_{0+}=v_{+-}=v_{-0}=\frac{1}{3}\left(\mathcal{V}_{1}+\mathcal{V}_{2}e^{i2\pi/3}+\mathcal{V}_{3}e^{-i2\pi/3}\right)=\frac{1}{3}\left(\mathcal{V}_{1}-\frac{1}{2}(\mathcal{V}_{2}+\mathcal{V}_{3})+i\frac{\sqrt{3}}{2}(\mathcal{V}_{2}-\mathcal{V}_{3})\right); \label{eq:offdia1}
\end{equation}

\begin{equation}
	v_{0-}=v_{-+}=v_{+0}=\frac{1}{3}\left(\mathcal{V}_{1}+\mathcal{V}_{2}e^{-i2\pi/3}+\mathcal{V}_{3}e^{i2\pi/3}\right)=\frac{1}{3}\left(\mathcal{V}_{1}-\frac{1}{2}(\mathcal{V}_{2}+\mathcal{V}_{3})-i\frac{\sqrt{3}}{2}(\mathcal{V}_{2}-\mathcal{V}_{3})\right). \label{eq:offdia2}
\end{equation}

Assuming $j=0$, and replacing the coefficients above  into Eq.     \ref{LSE-implicit-take-4-1D} we get:

\begin{equation}
\Psi_{0}=\frac{1}{\sqrt{2\pi\hbar}}-\frac{i}{2\hbar}\sqrt{\frac{2m}{E-\mathcal{E}_{0}}}\left(v_{0,0}\Psi_{0}+v_{0,+}\Psi_{+}+v_{0,-}\Psi_{-}\right),
\end{equation}

\begin{equation}
	\Psi_{+}=-\frac{i}{2\hbar}\sqrt{\frac{2m}{E-\mathcal{E}_{+}}}\left(v_{+,0}\Psi_{0}+v_{+,+}\Psi_{+}+v_{+,-}\Psi_{-}\right),
\end{equation}

\begin{equation}
\Psi_{-}=-\frac{i}{2\hbar}\sqrt{\frac{2m}{E-\mathcal{E}_{-}}}\left(v_{-,0}\Psi_{0}+v_{-,+}\Psi_{+}+v_{-,-}\Psi_{-}\right),
\end{equation}

 Solving the equations above and using \eqref{eq:dia},\eqref{eq:offdia1} and \eqref{eq:offdia2}, we get:

\begin{equation}
\Psi_{0}=N_\Psi\left[(-i (-i + b_- v_{0,0}) (-i + b_+ v_{0,0}) + i b_- b_+ v_{-,0} v_{+,0})\right],
\end{equation}

\begin{equation}
	\Psi_{+}=N_{\Psi}\left[b_+ (v_{+,0} - i b_- (v_{-,0}^2 - v_{0,0} v_{+,0})))\right],
	\end{equation}

\begin{equation}
\Psi_{-}=N_{\Psi}\left[b_- (v_{-,0} - i b_+(v_{+,0}^2-v_{0,0} v_{-,0}))\right],
\end{equation}

where $b_i=\frac{1}{2\hbar}\sqrt{\frac{2m}{E-\mathcal{E}_{i}}}$ and

\begin{gather}
\frac{1}{N_{\Psi}}=-i (-i + b_- v_{0,0}) (-i + b_+ v_{0,0}) + i b_- b_+ v_{-,0} v_{+,0} +  \notag \\
 b_0 \left(-i (b_- + b_+) v_{0,0}^2 + b_- b_+ v_{0,0}^3 +
 b_- b_+ v_{-,0}^3 + 
    i (b_- + b_+) v_{-,0} v_{+,0} + b_- b_+ v_{+,0}^3 - v_{0,0} (1 + 3 b_- b_+ v_{-,0} v_{+,0})\right).       
\end{gather}

In  a similar way, the scattering wavefunctions can be deduced for $j=+$ and $j=-$. Having the scattering wavefunctions it is straigthforward to obtain the $T$-Matrix elements using Eq. \ref{T-matels-take-3-1D} of this supplement.

\subsection{Detailed balance conditions on the T-matrix for a toy model}

In this section we show that our toy model breaks the four detailed balance conditions (see Section III on the main text).  In particular we will show that the necessary conditions for  $I(+,0)\neq1 $ are met. For our toy model:

\begin{equation}
    \langle p'+|T|p \, 0\rangle=\frac{iN_{\Psi}}{\sqrt{2\pi\hbar}}(v_{+,0}-ib_-(v_{-,0}^{2}-v_{0,0}v_{+,0})),
\end{equation}
\begin{equation}
\langle p \, 0|T|p'+\rangle=\frac{iN_{\Psi}}{\sqrt{2\pi\hbar}}(v_{-,0}-ib_-(v_{+,0}^{2}-v_{0,0}v_{-,0})).
\end{equation}
Here the normalization prefactor is given by
\begin{gather}
    \frac{1}{N_{\Psi}}=\notag\\
    i+b_0b_-b_+\mathcal{V}_1\mathcal{V}_2\mathcal{V}_3+\frac{1}{3}(b_0+b_-+b_+)\left(\mathcal{V}_1+\mathcal{V}_2+\mathcal{V}_3\right)-\frac{i}{3}\left(b_-b_++b_0(b_-+b_+)\right)\left( \mathcal{V}_1\mathcal{V}_2+\mathcal{V}_1\mathcal{V}_3+\mathcal{V}_2\mathcal{V}_3\right).
\end{gather}

Besides the resonances, $E=\mathcal{E}_j$, $N_{\Psi}$ is a finite number for finite potentials. This implies that $N_{\Psi}$ in general is nonzero.
\vspace{1cm}

\underline{\textit{Hermiticity}}

In order to check hermiticity we do a series expansion on the coupling strength, $v_{ij}$. Hermiticity is broken at second order, as shown below:

\begin{gather}
\langle p'+|T|p\,0\rangle-\langle p\,0|T|p'+\rangle^{*}
=\frac{-i}{\sqrt{2\pi\hbar}}
\left(2 Re \left(b_-\right)v_{-0}^2+2v_{+0}v_{00}\left(Re (b_+)+Re (b_0)\right)
\right)+O(v^3_{ij}),
\end{gather}

%\begin{gather}
%\langle p'+|T|p\,0\rangle-\langle p\,0|T|p'+\rangle^{*}
%=\frac{iN_{\Psi}b_-}{\sqrt{2\pi\hbar}}\left(\sqrt{3}(\mathcal{V}_2-\mathcal{V}_3)\mathcal{V}_1+i(\mathcal{V}_2(\mathcal{V}_1-\mathcal{V}_3)+\mathcal{V}_3(\mathcal{V}_1-\mathcal{V}_2))\right),
%\end{gather}

which  besides some very specific values of $E$ is different from zero as long as not all the potentials are the same.
\vspace{1cm}

\underline{\textit{Symmetric matrix}}

\begin{gather}
\langle p'+|T|p\,0\rangle-\langle p\,0|T|p'+\rangle
=\frac{N_{\Psi}(\mathcal{V}_{2}-\mathcal{V}_{3})}{\sqrt{6\pi\hbar}}\left(1+ib_{-}\mathcal{V}_{1}\right),
\end{gather}

which is different from zero as long as $\mathcal{V}_{2}\neq\mathcal{V}_{3}$. This expression is correct for any coupling strength.
\vspace{1cm}

\underline{\textit{Time reversal symmetry with and without parity transformation:}}

In our toy model the T-matrix is invariant under parity transformation. Therefore, these two conditions are actually the same and here we just prove the time reversal symmetry with parity transformation.

%\begin{equation}
 %   |\langle p'+|T|p\,0\rangle|^{2}-|\langle p\,0|T|p'+\rangle|^{2}=\frac{2iN_{\Psi}b_{-}}{3\sqrt{6\pi\hbar}}(\mathcal{V}_{1}-\mathcal{V}_{3})(\mathcal{V}_{2}-\mathcal{V}_{3})(\mathcal{V}_{1}-\mathcal{V}_{3}),
%\end{equation}

\begin{equation}
    |\langle p'+|T|p\,0\rangle|^{2}-|\langle p\,0|T|p'+\rangle|^{2}=\frac{|N_{\Psi}|^2Re(b_{-})}{3\sqrt{3}\pi\hbar}(\mathcal{V}_{1}-\mathcal{V}_{3})(\mathcal{V}_{2}-\mathcal{V}_{3})(\mathcal{V}_{1}-\mathcal{V}_{3}),
\end{equation}

which is different from zero as long as \emph{all} the potentials are different.   This expression is correct for any coupling strength.

\subsection{Calculating $I(k,l)$ for a toy model}

Despite breaking all the sufficient conditions on the T-matrix for DBE, there could be cases where DBE is still preserved. To confirm that DBE is actually broken, we explicitly calculate $I(k,l)$. For this, we numerically compute the following integrals:
      \begin{eqnarray}
      %  -------------------------------------------------------------------------------------------------------------------
         A_\beta\n(j'\m,j) & = & \int_{-\infty}^{+\infty} \m {\rm d}p \int_{-\infty}^{+\infty} \m {\rm d}p'\n \;\, e^{-\beta E_p} \;
         \delta\n\Bigl( E_p \, + \, {\cal E}_j \, - \, E_{p'\n} \, - \, {\cal E}_{j'\n} \Bigr) \;
         \Bigl| \, \la p'\m \,\n j'\n | \, T \, | p \,\n j \ra \, \Bigr|^2 \; = \nonumber\\
      %  -------------------------------------------------------------------------------------------------------------------
         & = & 2\,m \, \int_{{\cal E}_j}^\infty \m {\rm d}E \;\, \frac{e^{-\beta(E-{\cal E}_j)}}{\sqrt{(E-{\cal E}_j)(E-{\cal E}_{j'}\n)}} \;
         \Bigl| \, \la p'\m \,\n j'\n | \, T \, | p \,\n j \ra \, \Bigr|^2 ;
      %  -------------------------------------------------------------------------------------------------------------------
      \end{eqnarray}
      and
      \begin{eqnarray}
      %  -------------------------------------------------------------------------------------------------------------------
         B_\beta\n(j'\m,j) & = & \int_{-\infty}^{+\infty} \m {\rm d}p \int_{-\infty}^{+\infty} \m {\rm d}p'\n \;\, e^{-\beta E_p} \;
         \delta\n\Bigl( E_p \, + \, {\cal E}_j \, - \, E_{p'\n} \, - \, {\cal E}_{j'\n} \Bigr) \;
         \Bigl| \, \la p \,\n j | \, T \, | p'\m \,\n j'\n \ra \, \Bigr|^2 \; = \nonumber\\
      %  -------------------------------------------------------------------------------------------------------------------
         & = & 2\,m \, \int_{{\cal E}_j}^\infty \m {\rm d}E \;\, \frac{e^{-\beta(E-{\cal E}_j)}}{\sqrt{(E-{\cal E}_j)(E-{\cal E}_{j'}\n)}} \;
         \Bigl| \, \la p \,\n j | \,T \, | p'\m \,\n j'\n \ra \, \Bigr|^2 .
      %  -------------------------------------------------------------------------------------------------------------------
      \end{eqnarray}
      Subsequently, we also get the ratio
      \be
         I_\beta\n(j'\m,j) \; = \; \frac{A_\beta\n(j'\m,j)}{B_\beta\n(j'\m,j)}.
      \ee

      Having in hand $I_\beta\n(j'\m,j)$,
      we can even check the validity of the two thermalization conditions (Eqs. 30a and 30b in the main text).  The obtained numerical results are presented graphically in the main text.
%---------------------------------------------------------------------------------------------------------------------------
%

\end{document}